# Diagrammatic Modelling of Causality and Causal Relations

Sabah Al-Fedaghi*

*Computer Engineering Department*
*Kuwait University*
*Kuwait*

salfedaghi@yahoo.com, sabah.alfedaghi@ku.edu.kw

***Abstract* – It has been stated that the notion of *cause and effect* is one object of study that sciences and engineering revolve around. Lately, in software engineering, diagrammatic causal inference methods (e.g., Pearl's model) have gained popularity (e.g., analyzing causes and effects of change in software requirement development). This paper concerns diagrammatical (graphic) models of causal relationships. Specifically, we experiment with using the conceptual language of thinging machines (TMs) as a tool in this context. This would benefit works on causal relationships in requirements engineering, enhance our understanding of the TM modeling, and contribute to the study of the philosophical notion of causality. To specify the causality in a system's description is to constrain the system's behavior and thus exclude some possible chronologies of events. The notion of causality has been studied based on tools to express causal questions in diagrammatic and algebraic forms. Causal models deploy diagrammatic models, structural equations, and counterfactual and interventional logic. Diagrammatic models serve as a language for representing what we know about the world. The research methodology in the paper focuses on converting causal graphs into TM models and contrasts the two types of representation. The results show that the TM depiction of causality is more complete and therefore can provide a foundation for causal graphs.**

## I. INTRODUCTION

The graphical framework of the causal inference methods has gained popularity in software engineering [1]. Diagrammatic causal inference methods, such as Pearl's models, have been utilized in such areas as analyzing causes and effects of change in software requirement development [2]. Another usage is to estimate causal effects from observational data for explicit reasoning about issues related to correlations. Additionally, research fields are benefiting from the study of causal inference and reasoning in such areas as the formulation of search queries as well as inclusion and exclusion criteria of the search results [3].

Nevertheless, according to [3], most causal reasoning is done informally through an exploratory process of forming causality graphs. Causal reasoning is also used as a justification for many tools intended to make the software more human-readable by providing additional causal information to logging processes or modeling languages. Ref. [3] observed that causal inference is an up and coming field of study but is relatively underutilized in software engineering. Other applications are creating an evidence-based model, conducting observational studies on software fault localization, and developing causation-based strategies in the field of artificial intelligence for information technologies [3]. Ref. [4] mentioned the lack of explanatory and causality studies in software fault-proneness-related works on analysis and prediction.

### A. *Problems of Concern in this Paper*

#### 1) *"Mathematization" of Modeling:*

According to [5], *variables* are the basic building blocks of causal models. The values of a variable can represent the occurrence or nonoccurrence of an *event* or a range of incompatible events, For instance, a situation in which Suzy throws a stone and a window breaks may have variables $S$ and $W$ such that

$S = 1$ represents Suzy throwing a rock,
$S = 0$ represents her not throwing,
$W = 1$ represents the window breaking, and
$W = 0$ represents the window remaining intact.

But such a mathematical representation of reality looks like a shorthand hardware-oriented specification. It involves a trial-and-error procedure to map reality to variables based on mathematical patterns. This "mathematization" [6] reflects a drive for identifying *variables* as the cornerstone of solutions for nonmathematical situations or phenomena. If the solution does not accord with the reality, certain stages or the entire modeling process are repeated [6]. It "degrades" modelling to a definition as expressing real-life situations in mathematical language [6]. The abstractness of variables should motivate a more precise identification with their roots in reality.

We can conclude from such a discussion that interpreting real-life problems mathematically requires a clear conceptual representation of backgrounds. For example, the variables $S$ and $W$ given above fall under the usual definition of variables as *unknowns* that stand for a value that is not known yet. On the other hand, in the approach that we will discuss in this paper (simplified thinging machine (TM) modeling), the situation is presented as shown in Fig. 1.

---

*Retired June 2021, seconded fall semester 2021/2022

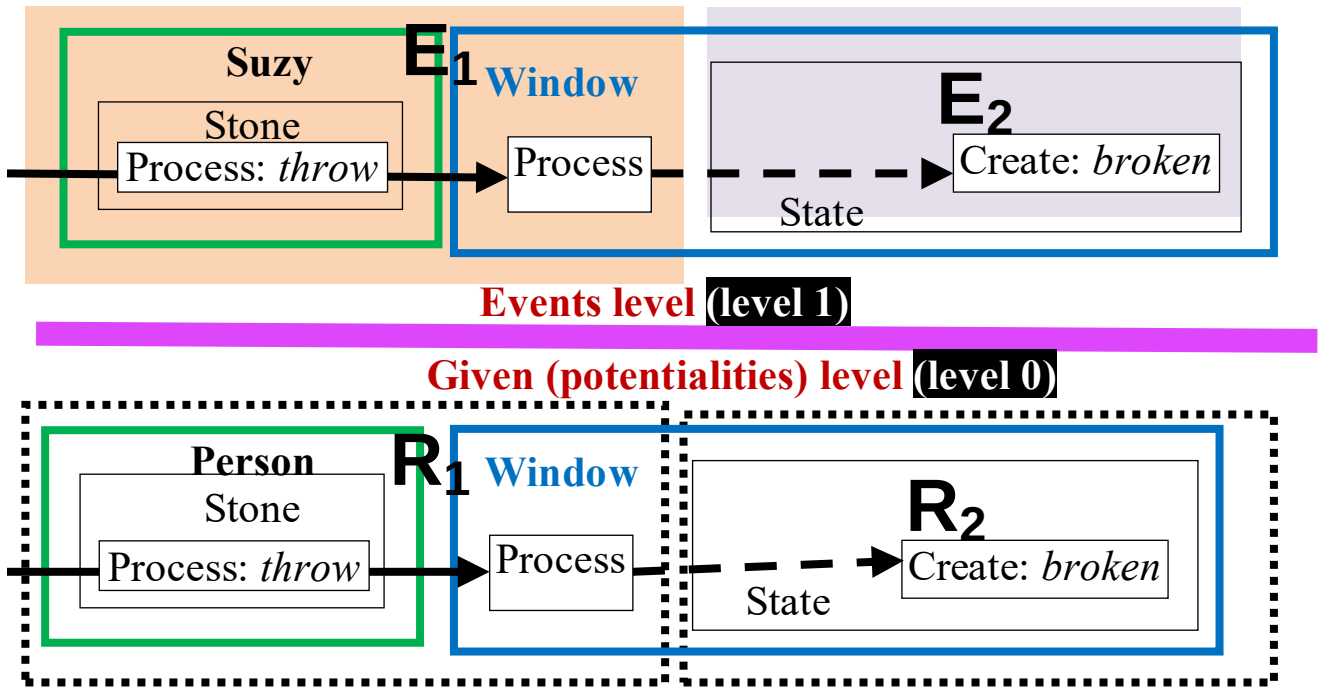

**Fig. 1 "Background" of *Suzy throwing a rock* and *window breaks*.**

In Fig. 1, the horizontal pink line separates what is given (static potential information) from the world of events. Therefore, $E_1$ and $E_2$ (colorbred boxes – level 1) are events and $R_1$ and $R_2$ (dotted boxes – level 0) are regions of events. The solid arrow in the figure denotes a flow (of stone), and the dashed arrows denote triggering.

$S = 1$ represents $E_1$, which is *Suzy throwing a rock*
$S = 0$ represents $R_1$, which is the absence of throwing
$W = 1$ represents $E_2$, which is the window breaking
$W = 0$ represents $R_2$, which is the absence of window breaking.

In such an approach, events are more meaningful than variables.

#### 2) *Difficulties in Causality*

Searching for cause is a seemingly unique human trait that dates back to (at least) Plato and Aristotle. The notion of causation played a central role in many philosophers' works (e.g., Descartes, Locke, Hume, and Kant). Currently, there are rich and extensive literatures in statistics, econometrics, and computer science on problems of causal inference and how best to understand the notion of causation [7]. It is even claimed that the object of study of all developments in sciences and engineering is a trio: matter (all forms of energy in all states), cause and effect relationships, and their behavior over time [8].

Still, a review of theoretical works in cognitive science suggests that causality remains ill understood [9]. According to [10], "causality is a notion shrouded in mystery, controversy, and caution, because scientists and philosophers have had difficulties defining when one event truly causes another." Inquiring in this context might demonstrate direct cause rather than broad correlation. *Correlation* occurs when one event is related to event but does not necessarily cause the other event to occur. *Causation* means one event causes another (cause and effect). Causation and correlation between events can exist simultaneously; however, correlation does not entail causation.

### B. *Improving the Conceptual Representation of Causality*

This paper focuses on the diagrammatic-modeling concept of TMs, which was developed originally in the context of software engineering [11-13]. A TM is a *model of the environment* (domain), in contrast to many other knowledge representation schemes (e.g., logic, rule-based systems, and neural networks) [14]. Such a TM application is meant to provide a better understanding of what "cause" actually is and may directly benefit a range of areas, such as reasoning, argumentation, learning, science comprehension, and communication.

In software engineering, models are often diagrammatic (i.e., nodes with arrows between them), which denotes structures [1]. Graphs are a well-known, well-understood, and frequently used means to represent the structural or behavioral properties of a software system (e.g., entity-relationship diagrams). We propose application of the TM model to causal relationships as an alternative diagrammatic representation with potential results in this area. The paper is not about causal relationships; rather, it is about the diagrammatic representations in such an area. The paper presents an exploratory study for further investigation of the notion of causality in terms of TM modeling. To illustrate this, we will

Section II.*A*: Describe the TM model briefly.
Section II.*B*: Describe a representative causal relationship called *firing squad* in a diagrammatic model called a causal graph.
Section II.*C*: Describe the same *firing squad* problem using TM representation.

Hopefully, we will thus introduce TM modeling as a viable tool that can be used to explain the notion of causality (the how) in a system. After this illustrative example in Section 2, the remainder of this paper involves the following:

- Section Three: A more elaborate description of TM modeling that includes re-modeling a state machine of a laptop running on a battery.
- Section Four: After we provide supplementary knowledge of TM modeling, this section returns to the topic of causal methods by introducing the so-called deterministic *structural equation models*, given in terms of a model of a gas grill used to cook meat.
- Section Five: In this section, we further demonstrate TM modeling of complex situations.

## II. TMs Described Briefly, Example

### A. *TMs Described Briefly*

We briefly describe the TM reality ontologically to provide an intuitive ontological picture of a TM model [16]. TM modeling describes reality on two levels: a potentiality/actuality scheme adopting an idea that goes back to the Stoic modes of reality. Fig. 2 defines the categorical structure of TM modeling.

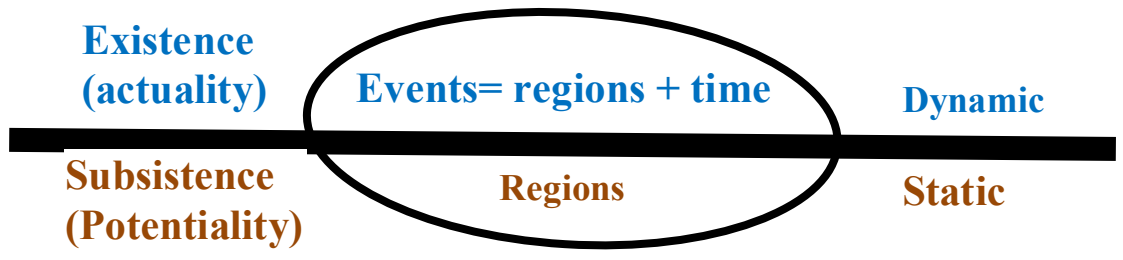

**Fig. 2 Two-level TM modeling.**

The basic thesis of this structure is

(1) Reality is a two-level system of static subsistence and dynamic existence.
(2) A region is a static description defined in terms of TMs (Fig. 3).
(3) A TM is a generalization of the known input-process-output model. The word *thinging* is a Heideggerian term, as explained in previous papers (e.g., [16]). The machine includes five actions: create, process (change), release, transfer, and receive.
(4) An event is defined in terms of region and time.

The general idea of this reality has been inspired by many thinkers. For example, according to [17],

> The virtual [potential in TM] is not opposed to the real but to the actual. The virtual is fully real in so far as it is virtual [potential].... Indeed, the virtual must be defined as strictly a part of the real [actual] object – as though the object had one part of itself [TM region] in the virtual into which it is plunged as though into an objective dimension.... The reality of the virtual consists of the differential elements and relations along with the singular points which correspond to them. The reality of the virtual is structure.

This type of ontology is centered on events. Accordingly, we claim that the *static* TM level represents the ontic (pre-categorical) reality. The two-level depiction is made to emphasize the characteristics of each the two levels. However, the TM model reflects two projected levels superimposed over each. An event and a region can exist simultaneously. Therefore, existence and subsistence are like a double-image impression (e.g., Rubin's vase), which is possible with a Gestaltic figure-ground perception. When we see an event, we simultaneously perceive its region. The region has real subsistence, but such a type of reality is "absently present" [18]. The mind can conceive quasi-real subsisting things purely in itself without considering their "existence," which is different from nonexistent (remember Rubin's vase, mentioned previously).

### *B. Sample Causal graph/diagram: The firing squad*

Cause and effect relationships can be visualized in many ways, including an cause and effect (Fishbone) diagram, causal graph, and cause and effect flowchart. We focus on causal diagrams.

Causal diagrams were developed to capture the activity in a real problem in preparation to put it in a mathematical/logical form. It is supposed to simulate the causal mechanisms (e.g., x "causes" y) that operates in the environment of the problem (the *domain* in software engineering). They also offer an intuitive approach to thinking about causal structures [19].

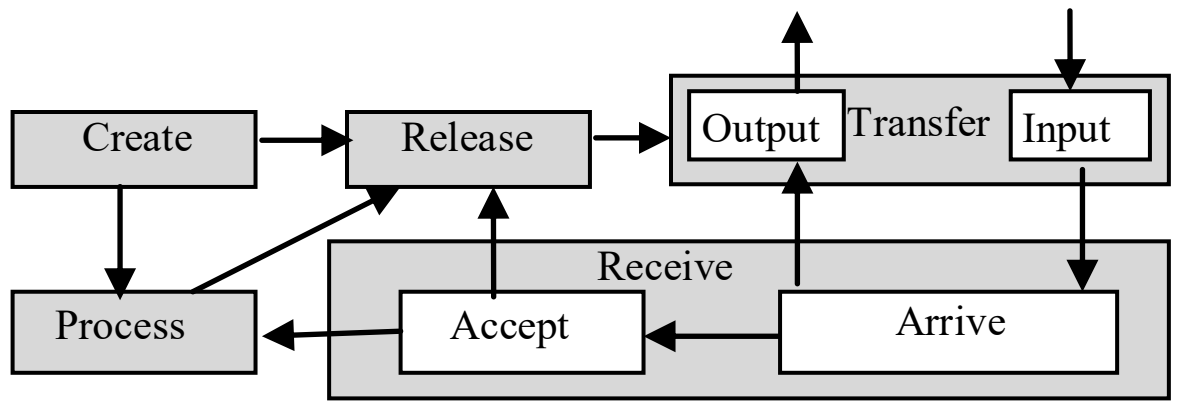

**Fig. 3 Thinging machine**

Consider the following example given by [20]. First, a court has to order the execution of a prisoner. The order goes to a captain, who signals the soldiers on the firing squad (A and B) to fire. We will assume that they are obedient and expert marksmen, so they only fire on command, and if either one of them shoots, the prisoner dies (see Fig. 4). In Fig. 4, each of the unknowns (CO, C, A, B, D) is a true/false variable. For example,

*D* = True means the prisoner is dead; *D* = False means the prisoner is alive.
*CO* = False means the court order was not issued; *CO* = True means it was, and so on [20].

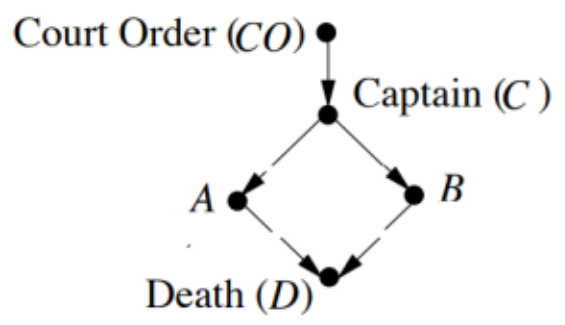

**Fig. 4 Causal diagram of the firing squad example. A and B represent (the actions of) soldiers A and B (adopted from [20]).**

### *C. TM Modeling of the Firing Squad*

TMs can be used to specify the notion of causality in a system's description to constrain the system's behavior and thus exclude some possible chronologies of events.

#### *1) Static TM Model of the Firing Squad*

TM modeling begins with building a static structural description that provides a base to develop a dynamic model that identifies events. Fig. 5 shows the static representation of the firing squad example.

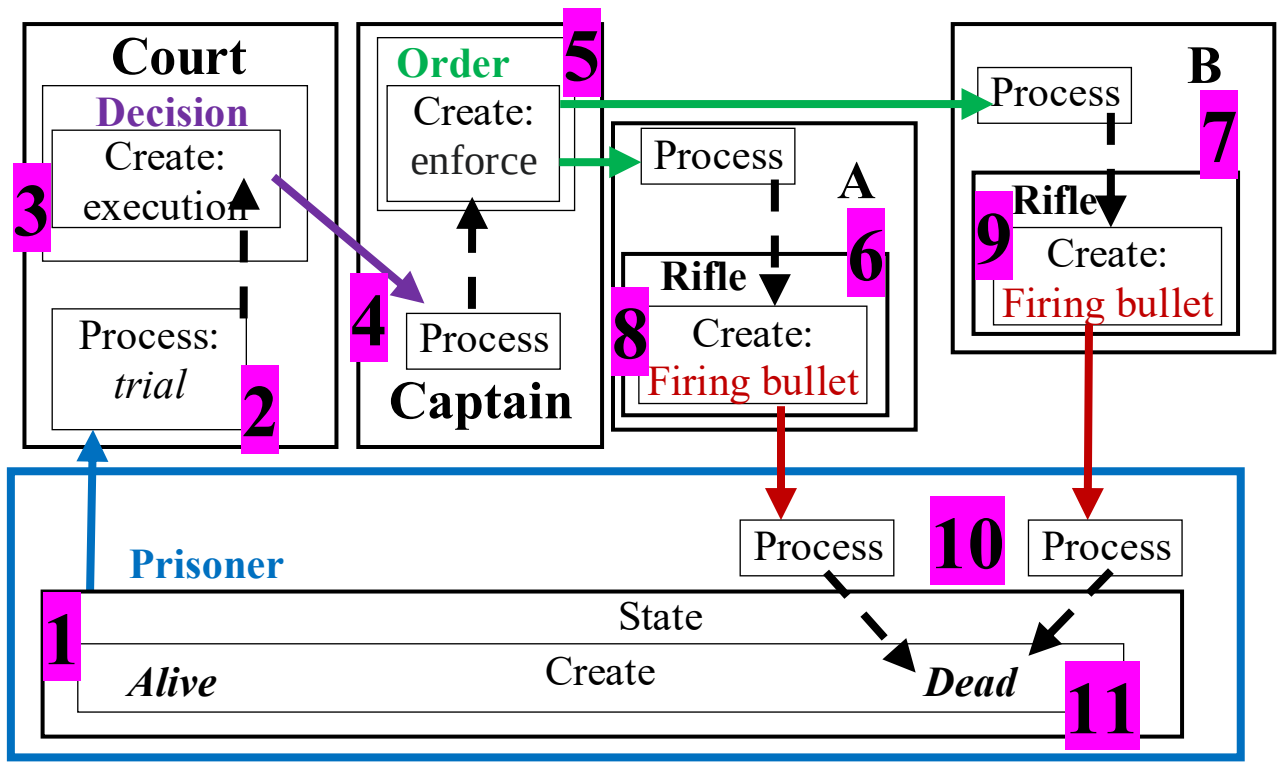

**Fig. 5 The static TM model of the firing squad example**

We made a simplification in Fig. 5 by eliminating the actions release, transfer, and receive based on the assumption that the directions of the arrows indicate the flow of things.

In Fig. 5, first, (pink number 1) a prisoner is tried in a court (2) and sentenced to death (3). The sentence is sent to the captain to be processed (4). The captain issues an order of execution (5) to soldiers A (6) and B (7). A and B fire (8 and 9) to hit (10) the prisoner, thus causing his/her death.

Notice that this static description is a map of potentialities of *processes* (in the common sense) that define the totality the firing squad procedure. It can be put in the form of a computer program that controls the procedure. It includes the steps of a plan that involves a prisoner, a court, a captain, soldiers, movement, communication, firing rifles, and "being-ness," such as life and death.

The total scenario is some*thing* over and above its constituents and a whole that is more than the sum of its parts. The totality of the scenario is some*thing* that is *real*, but it has *no temporal state* of nature and is not capable of causing effects in other things. Its reality is potential, static or has, in Stoic language, a subsistence nature. The firing squad procedure is not made up of material things (may *exist* in natural-language text, legal language in the form of law, or a computer program) or irreducible to the constituents from which it emerges (e.g., the fire squad issue could be a social, political, etc. issue; it may involve a supervisor, budget, and legal status). Additionally, parts of the procedure may be connected by flows (e.g., the prisoner goes to court, the court sends the captain a letter, etc.) or logical connections (e.g., the trial triggers the creation of the decision, rifle firing triggers the prisoner's death).

### *2) Dynamic TM model*

The firing squad description has its own "life" when realized in *time* at the dynamic level in terms of events. As mentioned previously, an event is defined in terms of region and time. For example, Fig. 6 shows the event *The prisoner goes to trial in court*. Note that the region is a subdiagram of the static-level diagram. For simplification, we represent events by their regions.

The event *The prisoner goes to trial in court* is a *connected event*. A connected event is an event with a connected region (connected subdiagram), not counting triggering. In this event, the variable *CO* = False means the court order was not issued, *CO* = True means it was, and so on [20]. *CO* is an abstract variable that ignores such details as the prisoner being alive. It is important also to note that *CO* ignores that *The prisoner **goes** to trial in court* triggered (causes/does not cause) the court order. This abstraction disregards the *initial* state of the prisoner and imbedded causal relation.

Accordingly, Fig. 7 shows the dynamic model of the firing squad example, in which regions are selected because they have connected events. The set of events is as follows:

$E_1$: *The prisoner is alive.* Note that this condition is an essential initialization because later, the prisoner will be dead. Additionally, note that the TM treats "entities" as events. This issue has been discussed in many papers (e.g., [16]).

$E_2$: *A prisoner is present in court for trial.*

$E_3$: *The court issues the decision to execute the prisoner.*

$E_4$: *The court order is sent to the captain.*

$E_5$: *The captain sends an order to soldiers A and B to fire on the prisoner.*

$E_6$: *Soldiers A and B fire bullets that hit the prisoner.*

$E_7$: *The prisoner is dead.*

We will not argue about the completeness of the TM model compared to that of causal graphs (Fig. 4) because comparing the two figures side by side is sufficient. It is possible to convert the TM model to a causal graph by eliminating details and keeping the variable regions.

Fig. 8 shows the chronology of events in the model where the ellipse indicates simultaneity. The arrows form a sequence or chain of events. This modeling leads us to study the notion of causality in its various manifestations, which is not the topic of this paper. The purpose of this paper is to propose TM modeling as a foundation for such a venture.

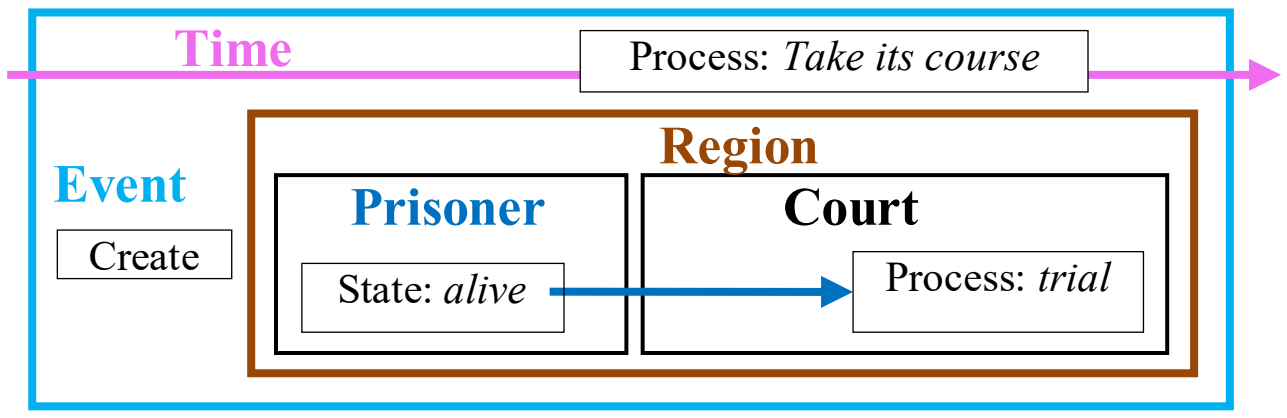

**Fig. 6 The event *The prisoner is tried in a court.***

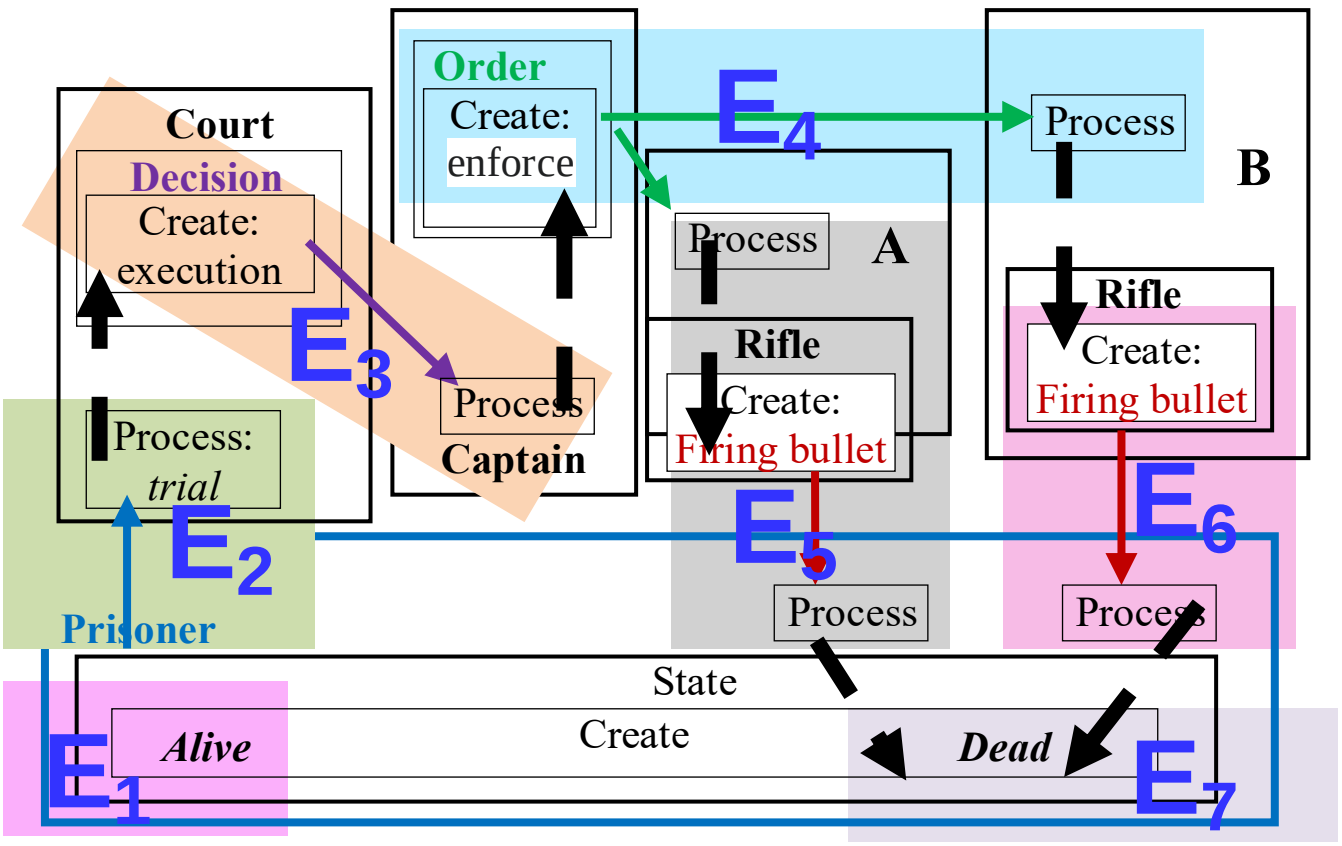

**Fig. 7 Connected events in the firing squad example**

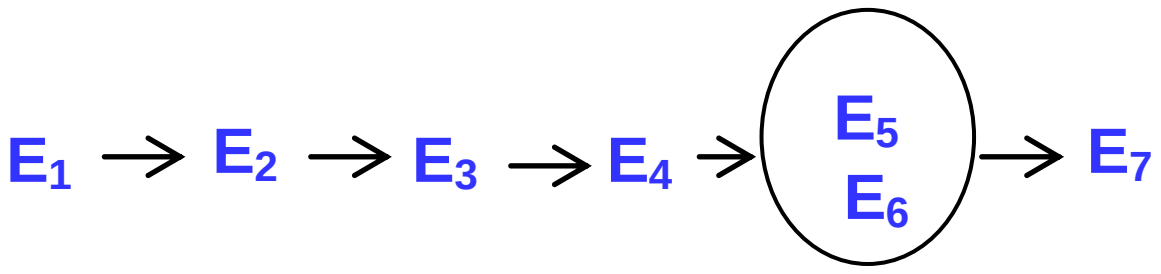

**Fig. 8 Chronology of events**

### D. Sample Analysis of Soldier A

According to [20], what if soldier A decides on his own initiative to fire, without waiting for the captain's command? Will the prisoner be dead or alive? Ref. [20] stated that this question has a *contradictory flavor* to it. Note that the new event (soldier A decides on his own initiative to fire) forms a subchain, $E_8 \rightarrow E_7$ (see Fig. 9), of the firing squad events, assuming the preservation of the chronology of events. According to [20], the solution means the removal of the edge from Captain (C) to A = True in the causal graph (Fig. 4).

From the TM point of view, such an analysis and solution are the result of the conceptualization of the causal graph representation. Such an analysis introduces a new potentiality (static level) and superimposes it on the original static description. Therefore, $E_8$ and $E_7$ (see Fig. 9) represent a new scenario that is overlaid over the old scenario. Fig. 10 shows the TM as a dynamic model resulting in a new chronology of events. Region $E_8$ coincides with part of region $E_6$.

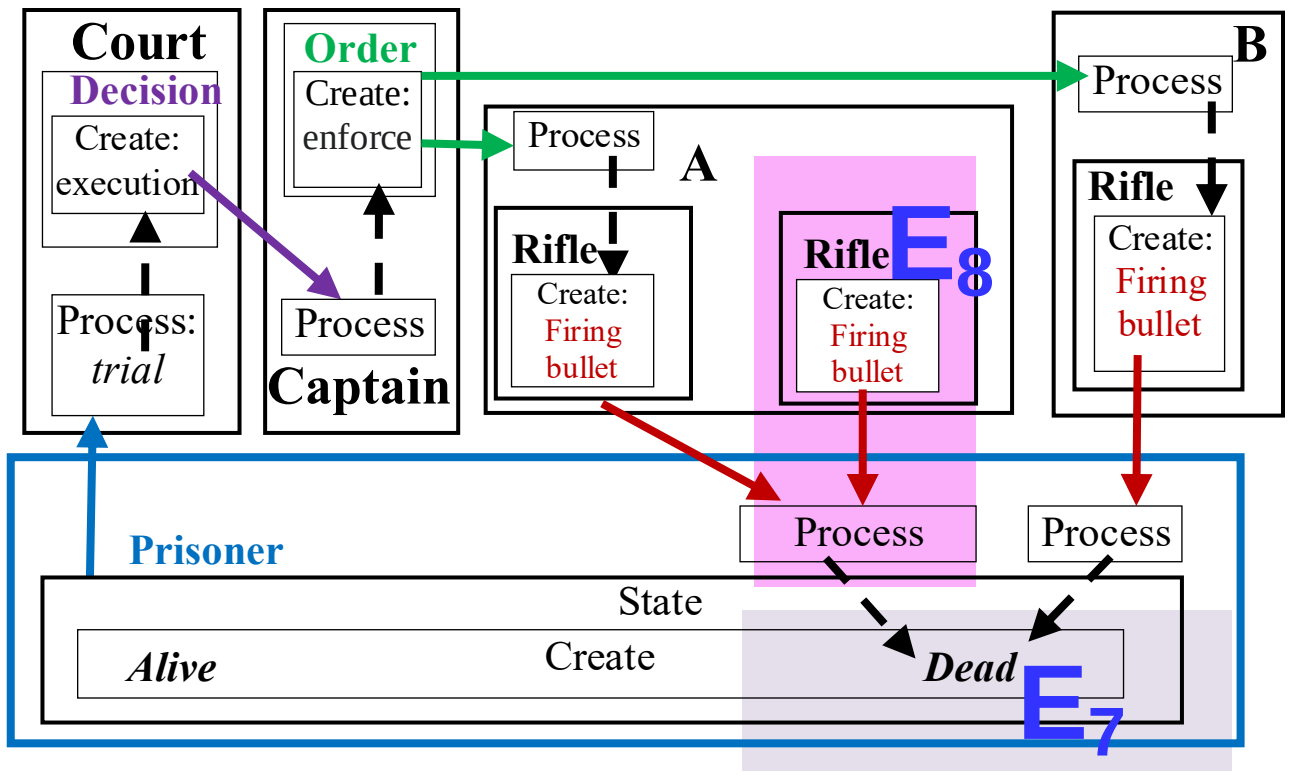

**Fig. 9 Adding a new event, A**

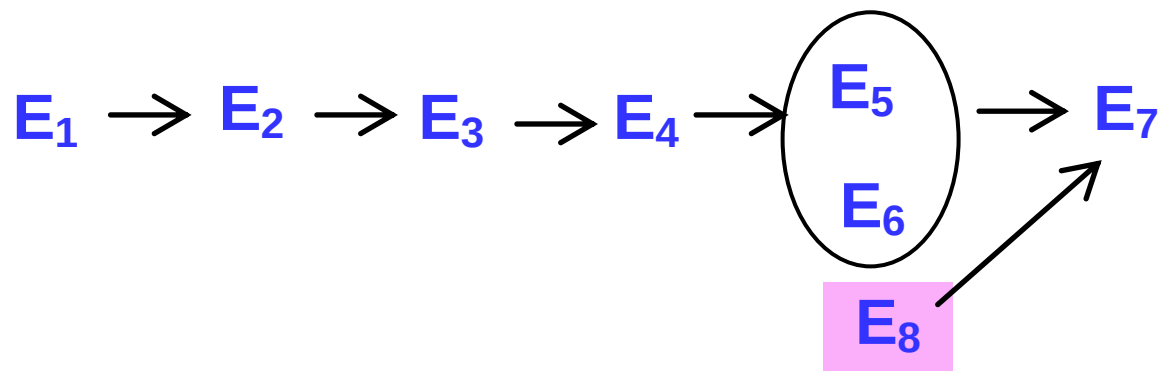

**Fig. 10 Adding event $E_8$ to the chronology of events**

## III. MORE ELABORATION ON THE TM MODEL

The TM model has been applied in many applications. We use each TM paper to develop the model further. Therefore, this section includes materials published previously and enhancement with new concepts. For example, special attention is paid to the notion of an event.

In TM modeling, "what is there?" is a world of *thimacs* (***thi**ngs/**mac**hines*) – a network of thimacs that articulate the furniture of the world. Thimacs are the only foundational elements of subsistence (atemporal) and existence (temporal) in reality. The world is made of composites of thimacs that interconnect with other thimacs. ***Hereafter, a thimac may be referred to as a thing or machine***. A thimac is a *machine* when it acts on other thimacs, and it is a *thing* when it is the object of actions by other thimacs*. A machine *things* (Heideggerian term); that is, it creates, processes (changes), receives, transfers, and releases.

### A. The Machine

In TM modeling, the thimac *machine* executes five actions: *create*, *process*, *release*, *transfer*, and *receive*. (See Fig. 3). Each of these static actions becomes a *generic event* when merged with time. Thimacs are realized by creating, processing, releasing, transferring and/or receiving thimacs.

The thimac *thing* is whatever is created, processed, released, transferred, and received. A thimac is a *machine* that creates, processes, releases, transfers, and receives. These actions are described as follows.

1) *Arrive:* A thing arrives to a machine.
1) *Accept:* A thing enters the machine. For simplification, we assume that arriving things are *accepted* (see Fig. 3); therefore, we can combine the *arrive* and *accept* actions into the *receive* action.
2) *Release:* A thing is ready for transfer outside the machine.
3) *Process:* A thing is changed, handled, and examined, but it is not transformed into a new thing.
4) *Transfer:* A thing is input into or output from a machine.
6) *Create:* A new thing is manifested in a machine.

In a TM, "create" has two meaning: (becoming) realized and (being) real. An example of the first sense is a thing coming into existence as the result of some processing of other things (emergence). In the second sense, a thing is declared an element of the domain's "inventory."

Additionally, the TM model includes a *triggering* mechanism (denoted by a **dashed arrow** in this article's figures), which initiates a (nonsequential) flow. Moreover, each action may have its own storage (denoted by a cylinder in the TM diagram). For simplicity, we may omit *create* from some diagrams because the box representing the thimac implies its being-ness (in the model). Additionally, the surrounding box of a machine may be omitted.

### B. Two-level modeling

The TM involves two *vertical* representations of a thing over a single model. A TM language assembles a model that has vertical dynamic representation over static representation. *Staticity* refers to *timelessness*. The static TM model is built from subsisting *regions* with a logical order imposed by *potential* flows and triggering. The static model comprises fixed parts, and it simply *subsists.*

Note that in this TM, dynamism and staticity are different from the similarly named notions. Consider the so-called *dynamic* events, such as *John walking*, and *static* events, such as *John resting under a tree* [21]. In a TM, both of these expressions are static regions that are actualized when merged with time to form events. Additionally, it is claimed that *walking* is a dynamic state, as opposed to a state of *resting*, which is static [21].

Next, we present a complete example of TM modeling to demonstrate the TM diagrammatic representation.

### C. *Example*

In The Unified Modeling Language, *states* represent the behavior of an object. A state machine models the lifetime of a single object. A state can contain other states, and in the graphical representation, the labeled edges depict transitions between states. Events are used to model occurrences (e.g., at an ATM machine, a user pressing a button to start a transaction) that trigger state transitions. An event (a signal, call, passing of time, or change in state) is a significant occurrence that has a location in time and space [22].

Reference [22] gives an interesting example of a state machine that involves a laptop running on a battery (See Fig. 11). When the laptop is on and the user is working, the laptop stays in the On state. However, after 5 minutes of user inactivity, the laptop turns off the screen to save power. Two transitions lead to the exit from the On state: a 5-minute timeout transition and a loop transition triggered by a keystroke or mouse movement. If the user touches the keyboard or mouse when the screen is off, the screen turns back on. If the user remains inactive for 10 minutes more, the laptop switches to sleep mode.

1) ***Static TM Model:*** Fig. 12 shows the corresponding static TM model.

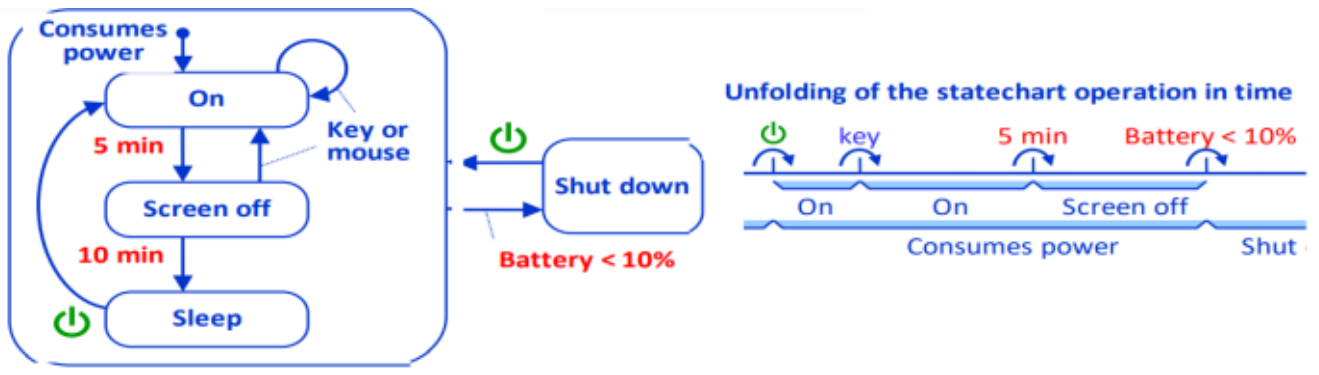

**Fig. 11 State chart of laptop running on battery (Partial from [22])**

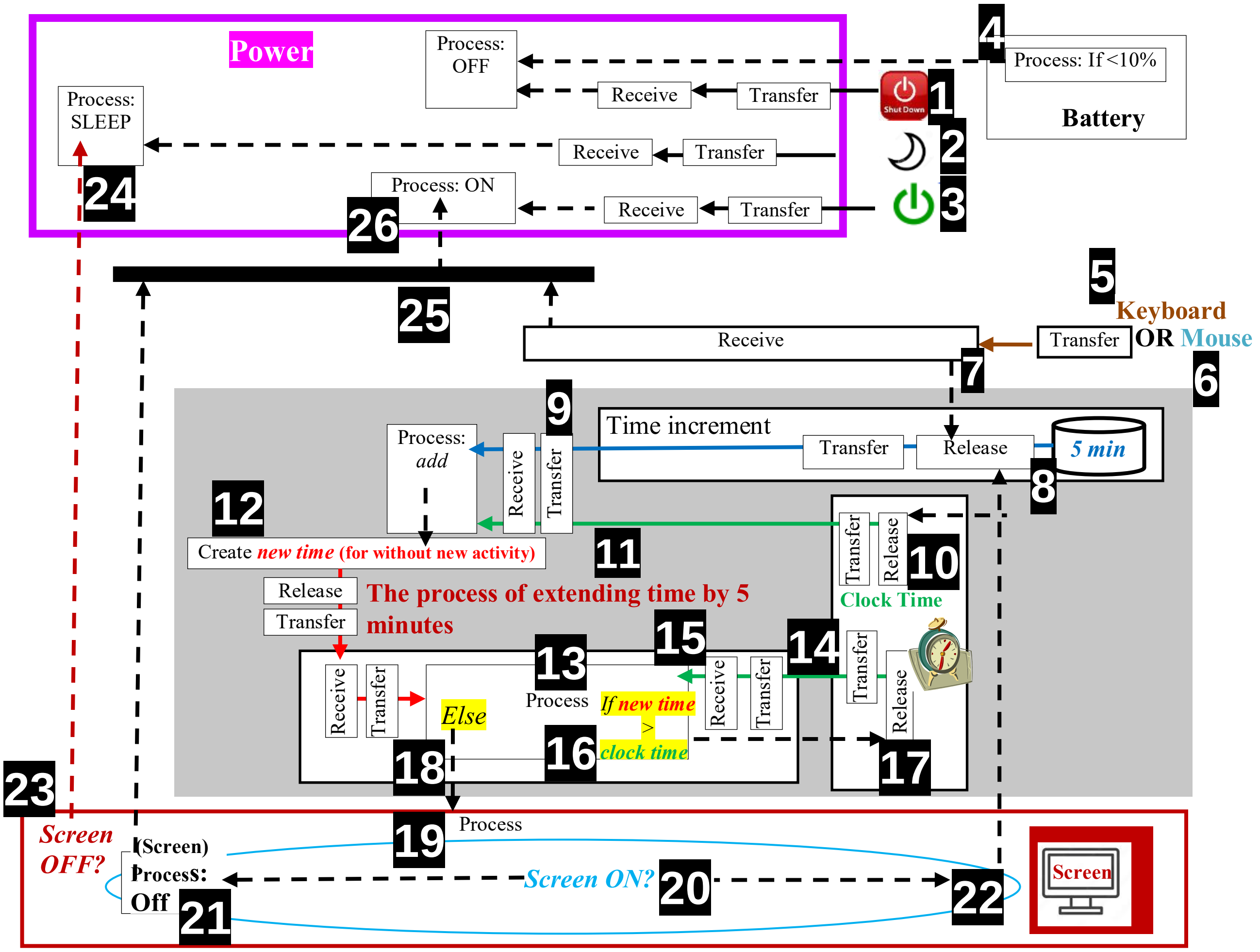

**Fig. 12 Static TM model**

The dark area in the figure controls the 5-minute increments. Numbers 1, 2, and 3 in the figure trigger power Shutdown, Sleep and power On (including screen and other components), respectively. If the battery power is <10%, the power turns Off (4).

Assuming that the power is On (3), the computer waits 5 minutes for further action. Then, working on the keyboard (5) or the mouse (6) triggers (7), retrieving the value "5" (8) to be added to the current time (9, 10, and 11) to create (12) the increment (time + 5 minutes) that represents starting the permitted non-activity period.

This new time (red arrow) is repeatedly compared (13) with the current time (green arrow, 14 and 15).

- *If new time > clock time* (16), the comparison is repeated (17) with the increasing current time.
- Otherwise (18), the screen is processed (19). If it is On (20), it is turned Off (21) and the process of 5-minute increments (dark area in the figure) continues (22). The next time, when the increment of time is 10 minutes (being inactive 10 minutes – Else 18) and the screen is Off (21), sleep is triggered (23 and 24).
- If working on the keyboard (5) or the mouse (6) **and** the screen is Off (23 and 24), the power is turned On (25 and 26). For simplification, the thick horizontal black line is used to the represent the "and" condition; otherwise, it can be represented in the TM language.

Fig. 12 can be simplified by eliminating the actions transfer, release, and receive under the assumption that the direction of the flow is represented by the direction of the arrows, as Fig. 13 shows.

2) ***Dynamic Model**:* Fig. 14 shows the dynamic model in terms of events. An event is defined as a region (subdiagram of the static model) and time. For example, Fig. 15 shows the event *Turning Power On*. For sake of simplicity, events are represented by their regions. Fig. 16 shows the chronology of events that form the cement that makes sense of the whole existing reality.

## IV. More on Causal Methods

Returning to the issue of causal relationships, [5] introduced the so-called deterministic *structural equation models* that characterize a causal system with variables and equations. Ref. [5] gives a model of a gas grill used to cook meat. The operations of the grill using the following variables are as follows:

- *Gas connected* (1 if yes, 0 if no)
- *Gas knob* (0 for off, 1 for low, 2 for medium, 3 for high)
- *Gas level* (0 for off, 1 for low, 2 for medium, 3 for high)
- *Igniter* (1 if pressed, 0 if not)

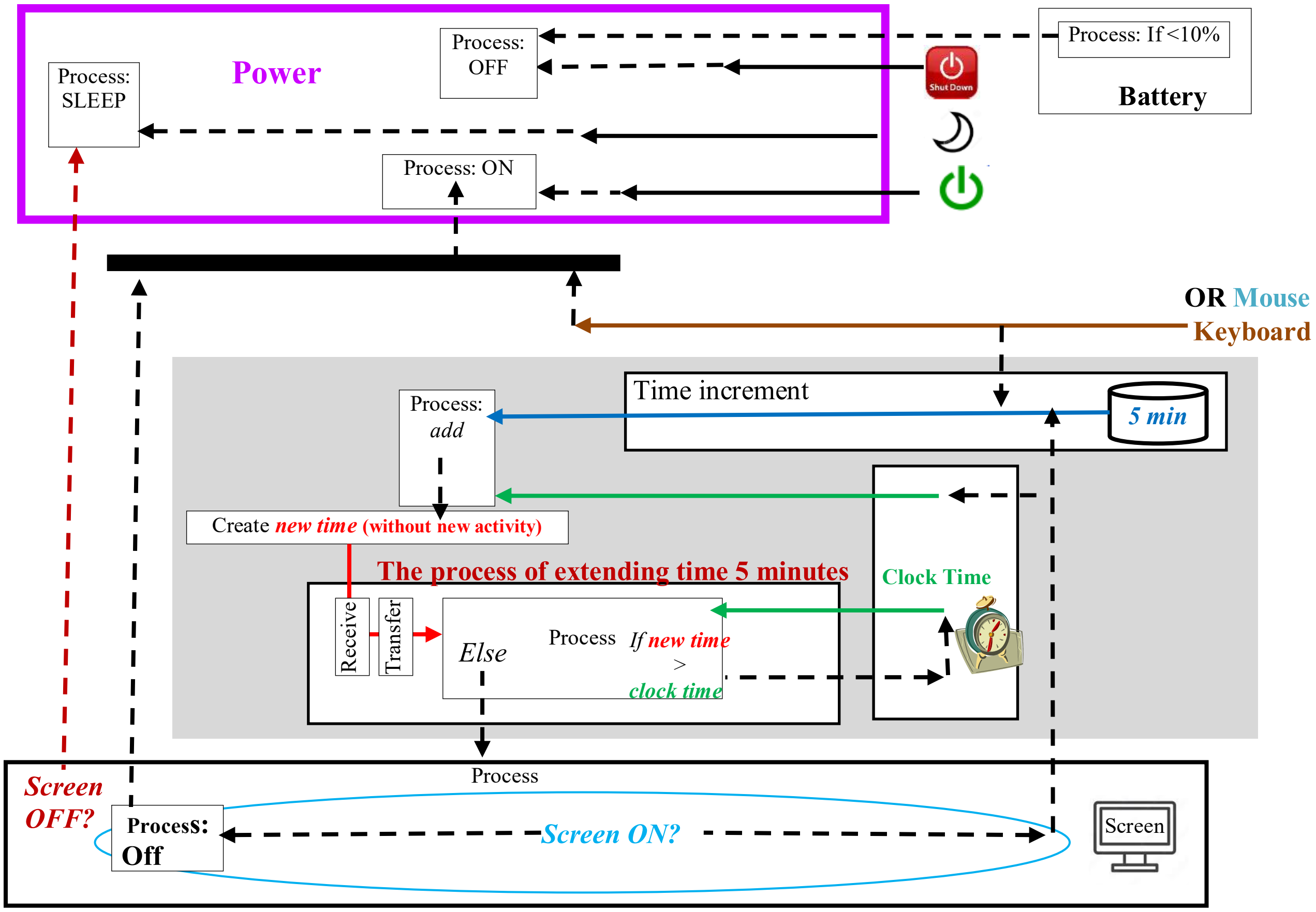

**Fig. 13 Simplification of Fig. 13, in which transfer, release, and receive are omitted**

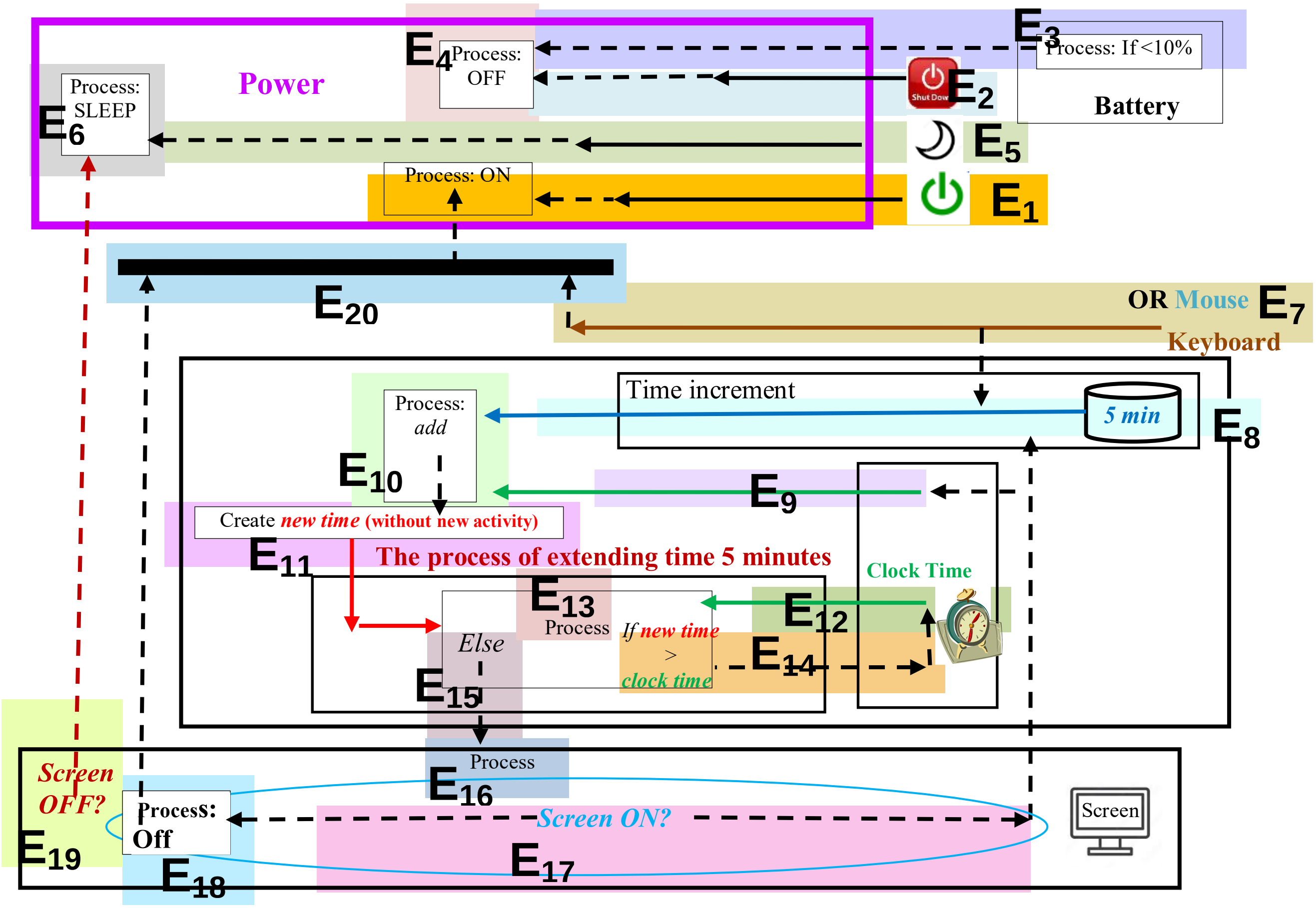

**Fig. 14 Dynamic model**

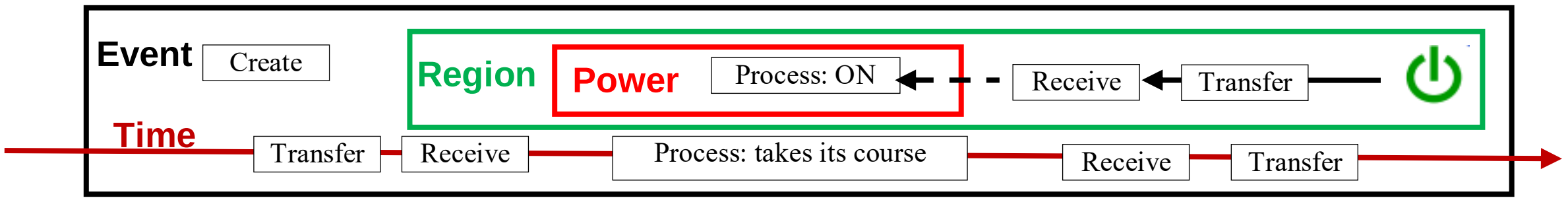

**Fig. 15 The event *Turning Power On***

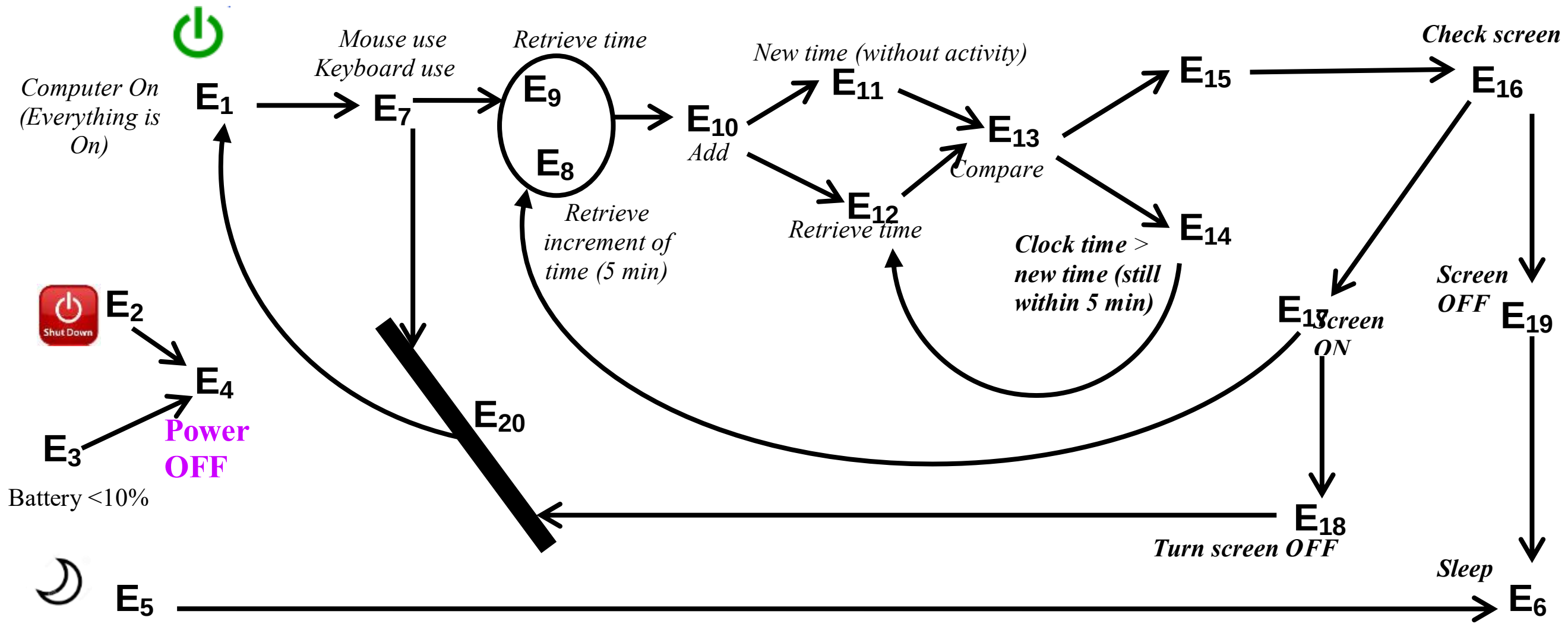

**Fig. 16 Chronology of events**

- *Flame* (0 for off, 1 for low, 2 for medium, 3 for high)
- *Meat on* (0 for no, 1 for yes)
- *Meat cooked* (0 for raw, 1 for rare, 2 for medium, 3 for well done)

Then the equations might be

- Gas level = Gas connected × Gas knob
- Flame = Gas level × Igniter
- Meat cooked = Flame × Meat on

The equations show that if the meat is not put on the grill, it will remain raw (*Meat cooked* = 0). If the meat is put on the grill, then it will be cooked according to the level of the flame: if the flame is low (*Flame* = 1), the meat will be rare (*Meat cooked* = 1) and so on [5]. Fig. 17 presents the system of equations as a graph.

Next, we will describe this structural equation model of a gas grill using a TM (See Fig. 18).

### A. *Static Model*

1. On the left side of Fig. 18 is the gas (1), which flows to the gas cooker (2). Inside that cooker, the knob has four levels to control the flow of the gas (4).
2. The gas is received (5) to be mixed with the spark the igniter generates (6) to create a flame (7 and 8).
3. The meat (9) is processed (10) by the flame to cook it, transforming it from raw meat to rare, medium, or well done (11).

### B. *Dynamic Model*

Fig. 19 shows the dynamic model of the gas grill. We select the following events:

$E_1$: The gas is connected.
$E_2$: Gas knob is set to 0 for off.
$E_3$: Gas knob is set 1 for low.

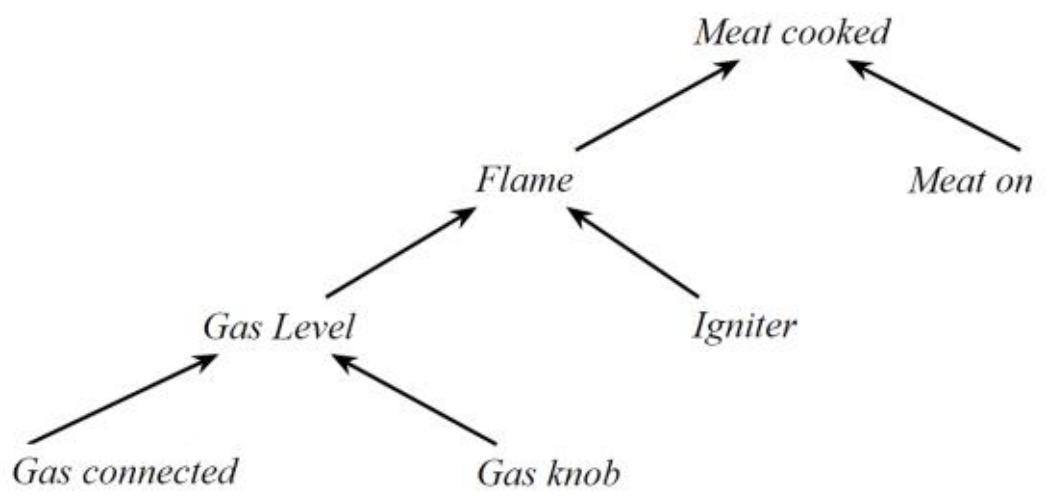

**Fig. 17 The representation of the system of equations (from [5])**

$E_4$: Gas knob is set to 2 for medium.
$E_5$: Gas knob is set to 3 for high.
$E_6$: The gas flows in to be mixed with a spark.
$E_7$: A spark is ignited and flows to be mixed with the gas.
$E_8$: Gas and spark are mixed.
$E_9$: Flame is generated.
$E_{10}$: Meat is received.
$E_{11}$: Meat is on the flame.
$E_{12}$: Meat is raw.
$E_{13}$: Meat is rare.
$E_{14}$: Meat is medium.
$E_{14}$: Meat is well done.

Note that we could have started with generic events that involve the five generic TM events: create, process, release, transfer, and receive. These events or sets of events are more complete for use in "propositions," as they are called in the structural equation models, to establish a logical level of inference.

In this paper, we focus on the representational base of the involved problem, leaving other issues, such as superimposing logic and probability treatment, for later research.

Fig. 20 shows the chronology of TM events and their mapping (dotted boxes) to the so-called variables of Fig. 17.

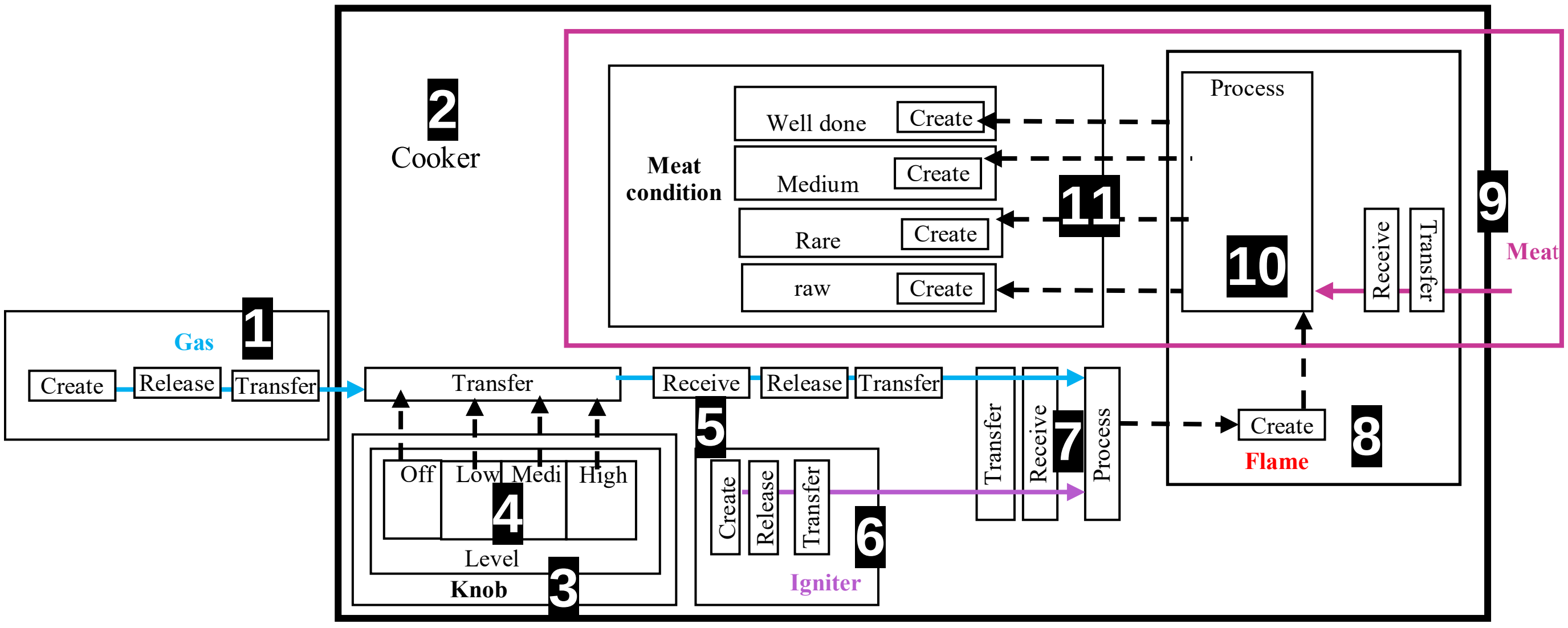

**Fig. 18 Static model**

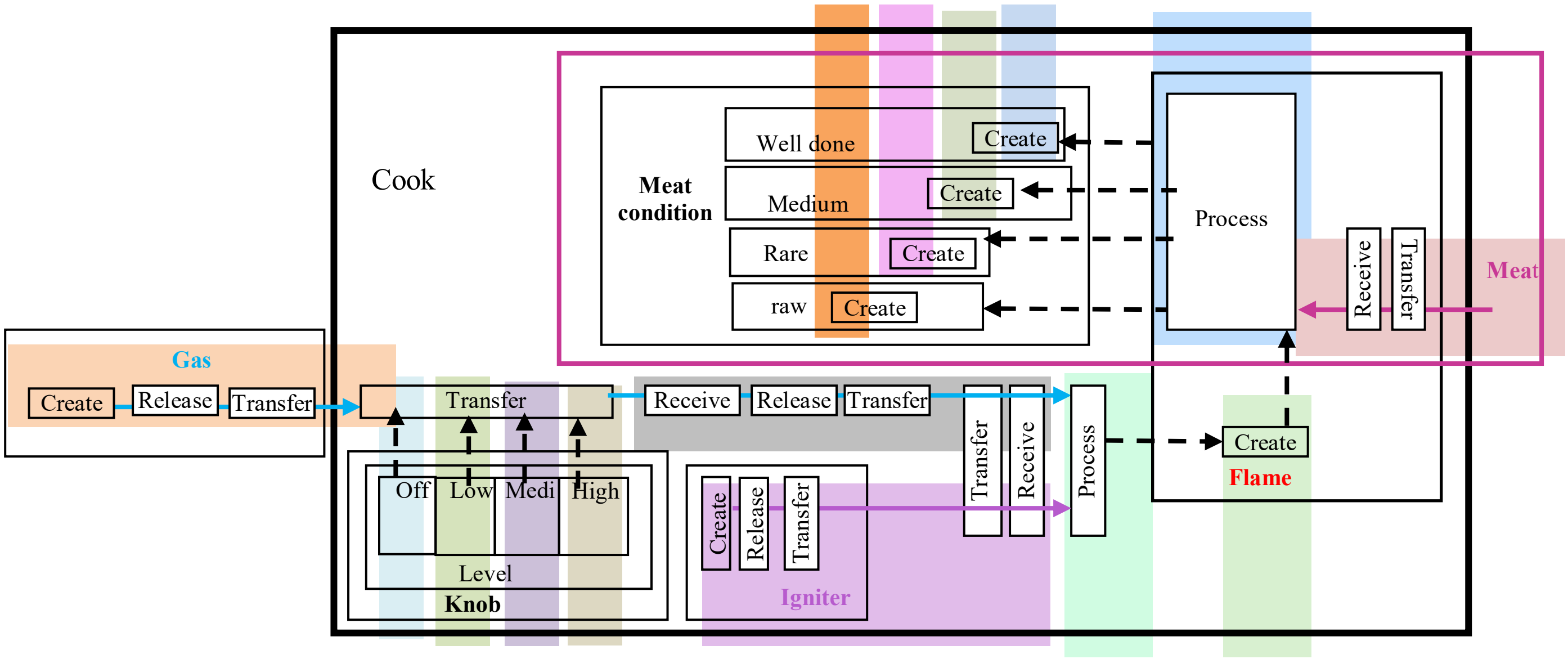

**Fig. 19 Dynamic model**

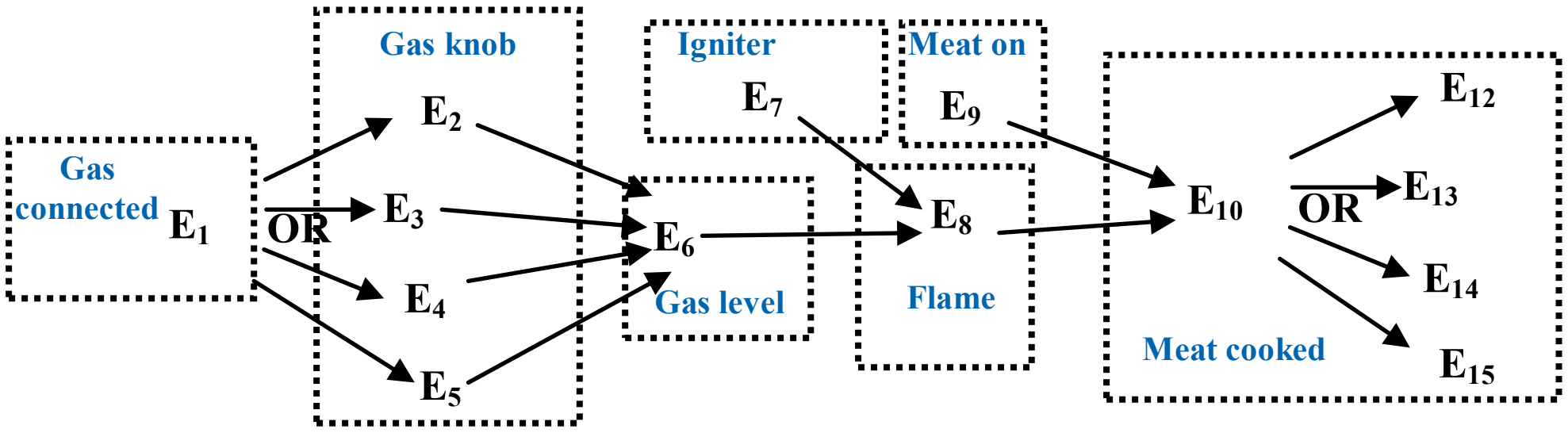

**Fig. 20 Chronology of events**

## C. *Sample Analysis*

To demonstrate one expressive feature of the TM representation, consider the following *actual causation* example that concerns the assignment of causal responsibility for some event that occurs based on how events actually play out [5].

Suppose that Billy and Suzy are holding rocks. Suzy throws her rock at a window, but Billy does not. Suzy's rock hits the window, which breaks. Suzy's throw was the cause of the window breaking. According to [5], we cannot assume the relation of actual causation from the graph or the equations.

We can represent this story with a TM diagram, as Fig. 21 shows. At the static (potential) level are *Person* (1) and *Another person* (2), and both have stones (3 and 4) that can be thrown and break the window. In actuality (the existence level), Suzy throws the stone and breaks the window ($E_1$ and $E_2$), but there is also Billy with his stone ($E_3$).

From the point of view of causal determinism as described in [23], events can exist or cease to exist without regions or events.

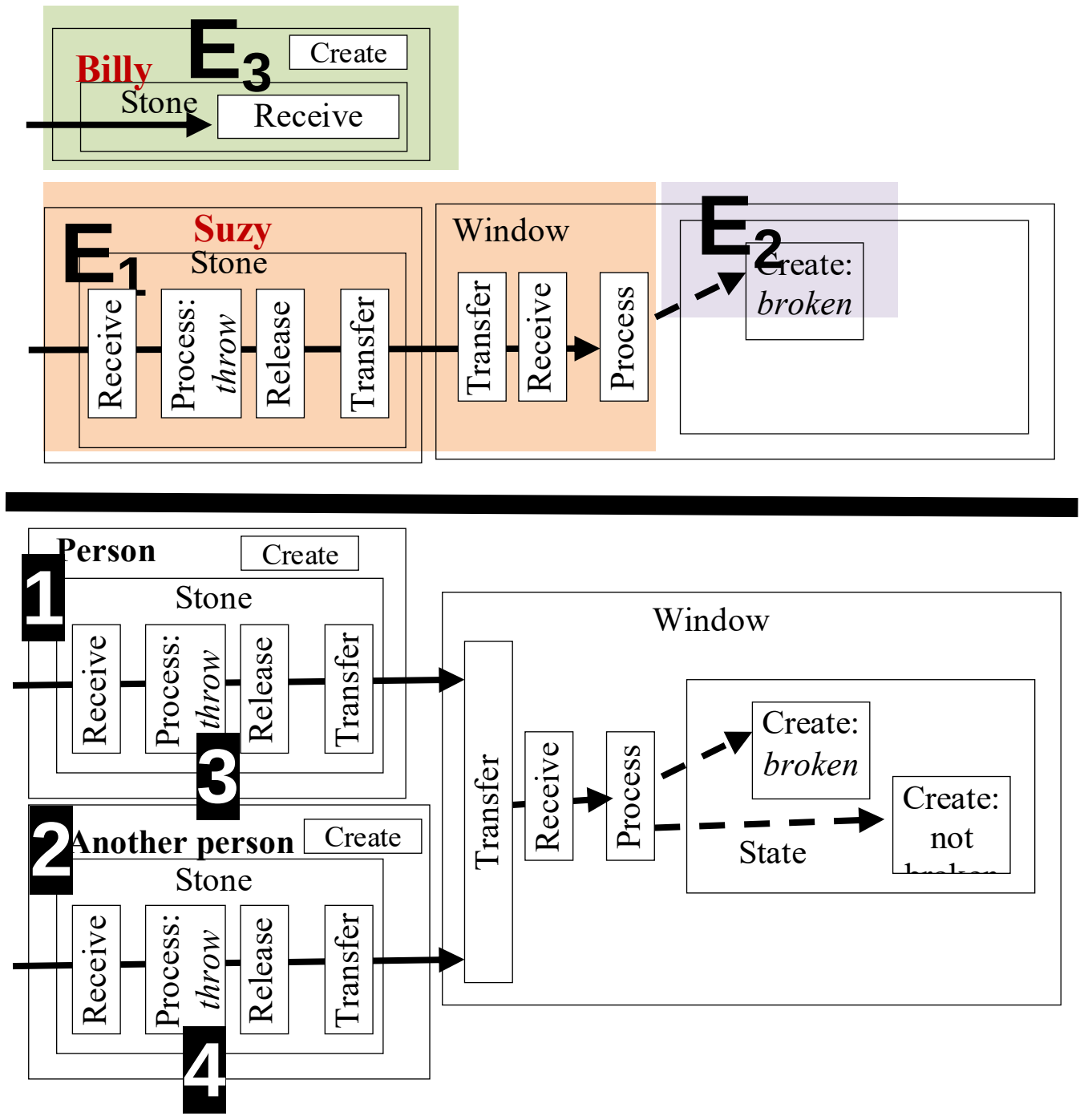

**Fig. 21 In actuality, Billy is there, but he did not throw his stone.**

In a TM, the process of realization involves a region that has become an event. Analogous to mathematical/logical equations and in a nonnumerical setting, we can specify *Event = R (region, time)*, where *R* is the process of realization. This indicates a region acquiring the property of event-ness. Therefore, for example, a *coin throwing* event involves a region of a coin and of a throwing agent, a horizontal surface, a gravitational field, and alternative yields of heads and tails. The adequacy of the description of the regions is essential for the correctness and consistency of mathematical/logical treatment. Therefore, when in the gas grill example such equations as *Gas level = Gas connected* × *Gas knob* and *Flame* = *Gas level* ×= *Igniter* are declared based on supposition and guesses without sufficient information, we expect an incomplete background even though experts may achieve success with small problems. On the other hand, TM modeling presents a more definite procedure as a base for moving to mathematical and logical treatment of the problem. It also helps improve elementary analyses of the causal relation in contrast to that which we get from natural language.

According to [23], in some studies of causality, researchers have examined the notion of causal relation in the context of ordinary language, occasionally adding "a pinch of logic." The latter often serves as "a carpet under which the conceptual muddles of common sense are swept" [23].

## V. OTHER ASPECTS OF CAUSALITY

In this section, we further demonstrate TM modeling of complex situations.

### *A. Properties and Capacities*

According to [24], a knife is partly defined by its properties (e.g., weight) and as being in a certain state, such as the state of being sharp. A sharp knife can cut things, an ability that can be exercised by interacting with entities that can be cut (e.g., meat). Philosophically, there is an important distinction between properties and abilities. Properties are always actual because at any given point in time, the knife is either sharp or it is not. But the *causal capacity* to cut is not necessarily actual if the knife is not currently being used. This implies that abilities can be *real* without being *actual.*

Fig. 22 shows the two-level representation of the knife system with two possible actualizations at the top level of existence. In the actualization that is immediately above the thick line, the knife is sharp and it cuts the meat. $E_1 \rightarrow E_2$ is a chronological and causal relationship. In the top actualization, the knife is not sharp; therefore, the meat is not cut. $E_3 \rightarrow E_4$ is not a causal relationship.

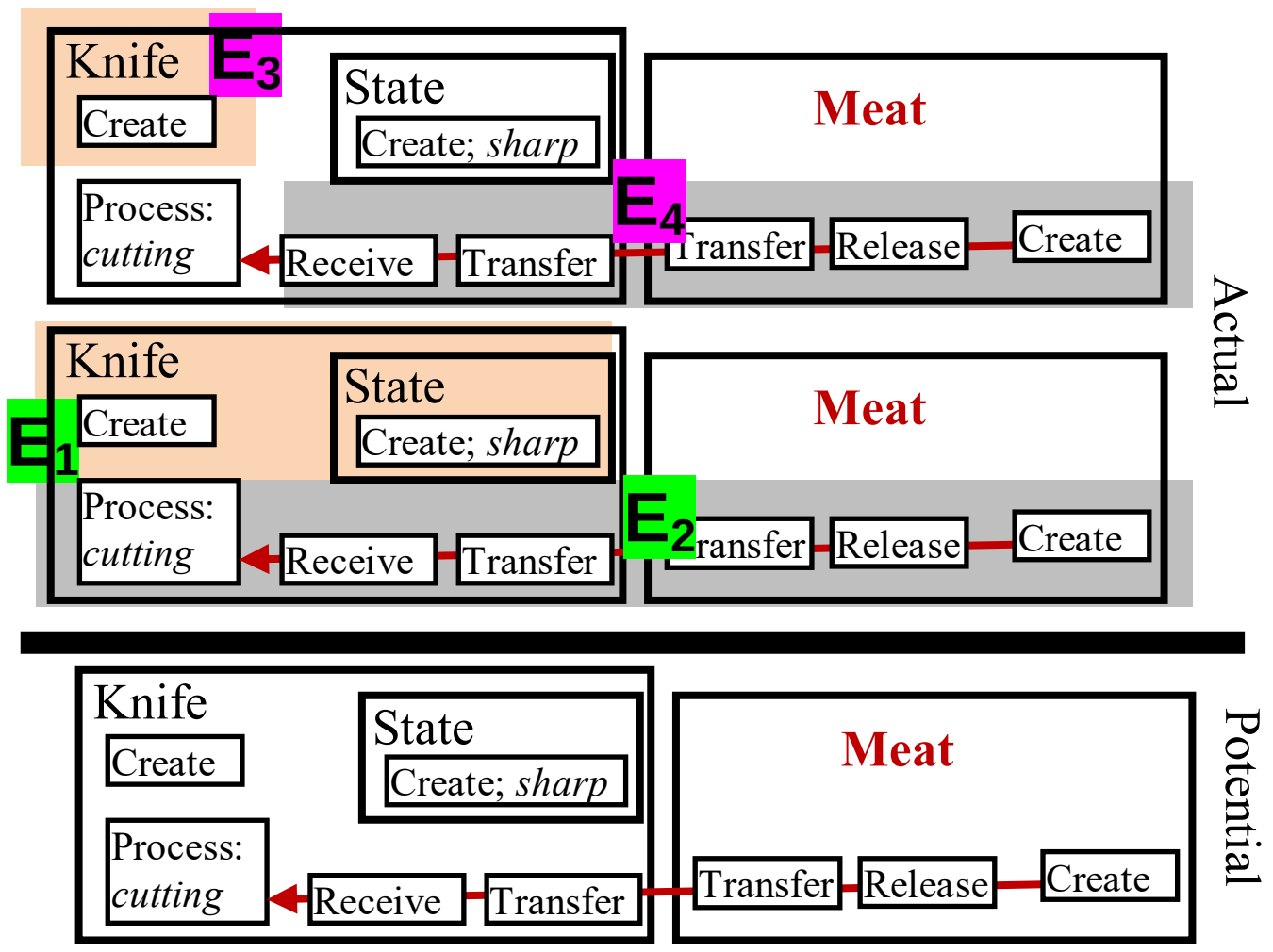

**Fig. 22 $E_1 \rightarrow E_2$ is a chronological and causal relationship and $E_3 \rightarrow E_4$ is not causal relationship.**

### *B. Negative Causation*

Many philosophers who have written on the subject reject the idea that "negative causation" is real causation [25]. Presumably, absences are just nothing, and nothings cannot be causally productive because "simply attempting to talk about 'nothings' as capable of being produced or being productive is to speak nonsense" [25]. The Franco-Romanian philosopher Stéphane Lupasco declared that every real phenomenon or event is always associated with an anti-phenomenon or anti-event such that the actualization of an event entails the potentialization of non-event and vice versa, alternatively, without either ever disappearing completely [26].

Consider the following example from [27]. Flora normally waters her neighbor's flowers, but she stops watering them, and they die. Common sense affirms that Flora's failure to water the plant is a cause of their death. Fig. 23 shows the TM representation, where *R* represents a negative event. In the figure, the repeated visiting and watering keeps the flowers in the state of being alive and growing. Repeated non-watering causes the state of being alive to cease.

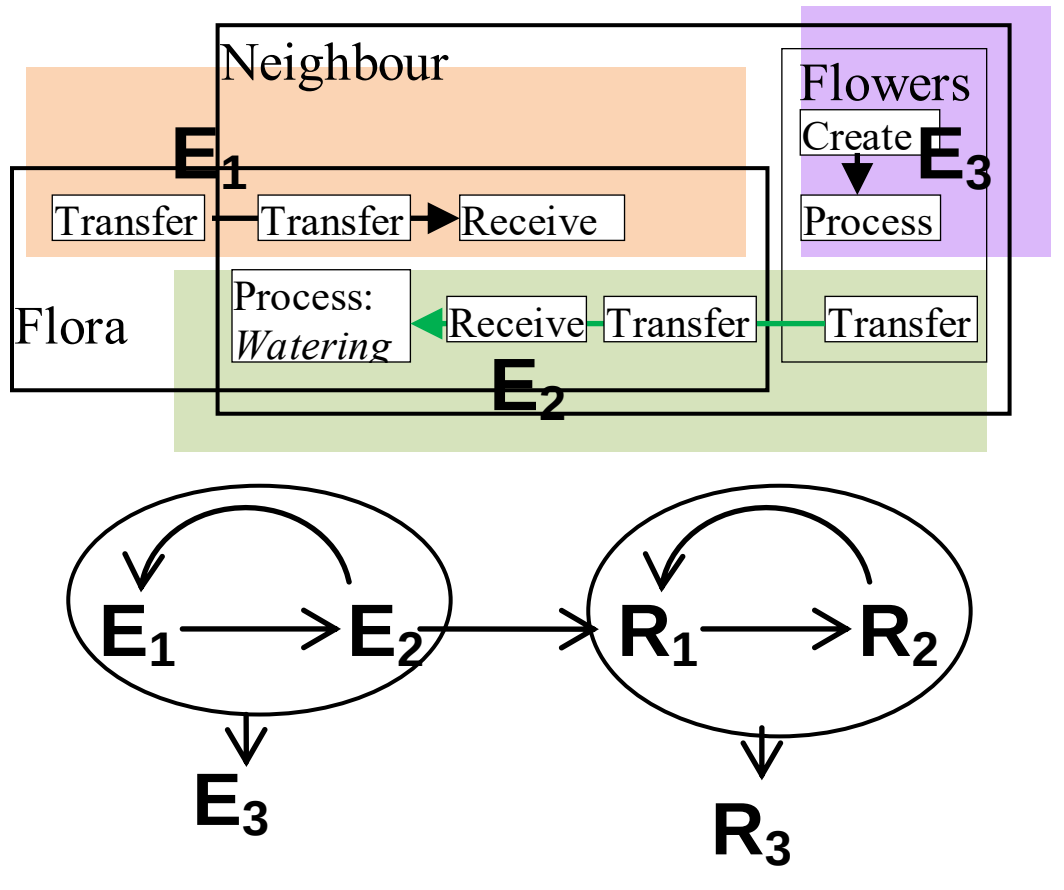

**Fig. 23 Flora's repeated watering of the flowers kept them alive; however, when she repeatedly fails to water them, they die.**

Consider the example from [27] (referring to Achille Varzi), *Johnny didn't turn off the gas because he got absorbed in his book.* Fig. 24 shows the TM representation. Repeatability is used to indicate divisibility of $E_2$; therefore, there are two options,

- Ending earlier and turning off the gas
- Taking a longer time and not turning the gas off

## VI. CONCLUSION

This paper concerns diagrammatical (graphic) models of causal relationships. Specifically, we experimented with using the conceptual language of TMs as a tool in this context. The many examples concerned TMs' suitability as a base for this purpose. However, the step of transforming the TM diagram to a logical or mathematical language needs further research.

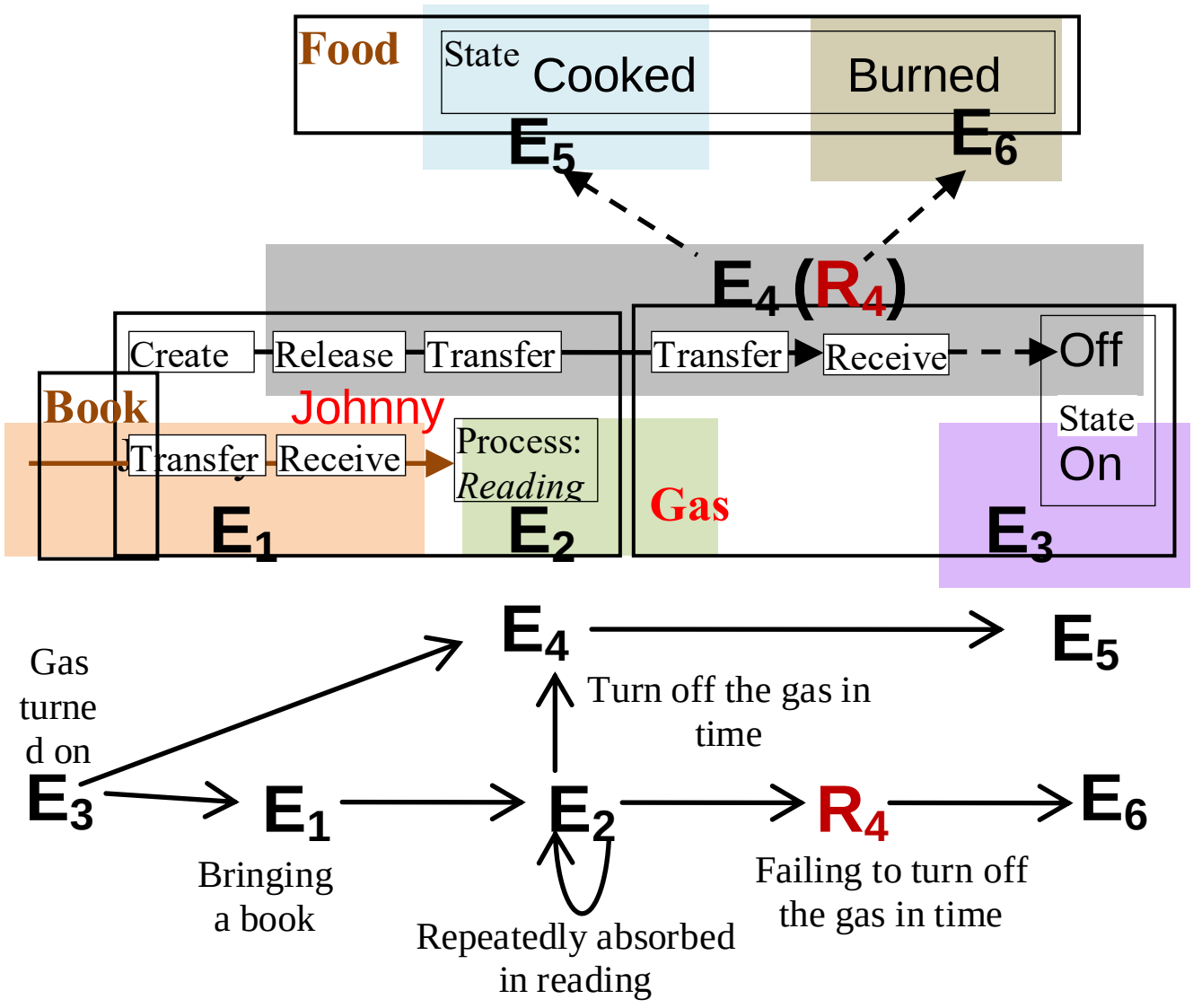

**Fig. 24 TM dynamic representation and chronology of events for *Johnny didn't turn off the gas because he got absorbed in his book.***